\documentclass[preprint] {aastex}

\usepackage{lineno}

\usepackage{epstopdf}
\usepackage{gensymb}
\usepackage{natbib}

\newcommand{\Fermic}{\textit{Fermi}}
\newcommand{\Fermi}{\Fermic\ }
\newcommand{\FermiLATc}{\Fermic-LAT}
\newcommand{\FermiLAT}{\FermiLATc\ }
\def\us{\char`\_}

\title{Spectral Energy Distribution of Markarian 501:  Quiescent State vs. Extreme Outburst}

\author{
V.~A.~Acciari\altaffilmark{1},
T.~Arlen\altaffilmark{2},
T.~Aune\altaffilmark{3},
M.~Beilicke\altaffilmark{4},
W.~Benbow\altaffilmark{1},
M.~B{\"o}ttcher\altaffilmark{5},
D.~Boltuch\altaffilmark{6},
S.~M.~Bradbury\altaffilmark{7},
J.~H.~Buckley\altaffilmark{4},
V.~Bugaev\altaffilmark{4},
A.~Cannon\altaffilmark{8},
A.~Cesarini\altaffilmark{9},
L.~Ciupik\altaffilmark{10},
W.~Cui\altaffilmark{11,\dagger},
R.~Dickherber\altaffilmark{4},
C.~Duke\altaffilmark{12},
M.~Errando\altaffilmark{13},
A.~Falcone\altaffilmark{14},
J.~P.~Finley\altaffilmark{11},
G.~Finnegan\altaffilmark{15},
L.~Fortson\altaffilmark{10},
A.~Furniss\altaffilmark{3},
N.~Galante\altaffilmark{1},
D.~Gall\altaffilmark{11,a,\dagger},
S.~Godambe\altaffilmark{15},
J.~Grube\altaffilmark{10},
R.~Guenette\altaffilmark{16},
G.~Gyuk\altaffilmark{10},
D.~Hanna\altaffilmark{16},
J.~Holder\altaffilmark{6},
D.~Huang\altaffilmark{17},
C.~M.~Hui\altaffilmark{15},
T.~B.~Humensky\altaffilmark{18},
A.~Imran\altaffilmark{19},
P.~Kaaret\altaffilmark{20},
N.~Karlsson\altaffilmark{10},
M.~Kertzman\altaffilmark{21},
D.~Kieda\altaffilmark{15},
A.~Konopelko\altaffilmark{17},
H.~Krawczynski\altaffilmark{4},
F.~Krennrich\altaffilmark{19},
A.~S~Madhavan\altaffilmark{19},
G.~Maier\altaffilmark{16,b},
S.~McArthur\altaffilmark{4},
A.~McCann\altaffilmark{16},
P.~Moriarty\altaffilmark{22},
R.~A.~Ong\altaffilmark{2},
A.~N.~Otte\altaffilmark{3},
D.~Pandel\altaffilmark{20},
J.~S.~Perkins\altaffilmark{1},
A.~Pichel\altaffilmark{23},
M.~Pohl\altaffilmark{19,c},
J.~Quinn\altaffilmark{8},
K.~Ragan\altaffilmark{16},
L.~C.~Reyes\altaffilmark{24},
P.~T.~Reynolds\altaffilmark{25},
E.~Roache\altaffilmark{1},
H.~J.~Rose\altaffilmark{7},
M.~Schroedter\altaffilmark{19},
G.~H.~Sembroski\altaffilmark{11},
D.~Steele\altaffilmark{10,d},
S.~P.~Swordy\altaffilmark{18},
M.~Theiling\altaffilmark{1},
S.~Thibadeau\altaffilmark{4},
A.~Varlotta\altaffilmark{11},
V.~V.~Vassiliev\altaffilmark{2},
S.~Vincent\altaffilmark{15},
S.~P.~Wakely\altaffilmark{18},
J.~E.~Ward\altaffilmark{8},
T.~C.~Weekes\altaffilmark{1},
A.~Weinstein\altaffilmark{2},
T.~Weisgarber\altaffilmark{18},
D.~A.~Williams\altaffilmark{3},
M.~Wood\altaffilmark{2},
B.~Zitzer\altaffilmark{11}
\newline
(the VERITAS Collaboration)
\newline
 J.~Aleksi\'c\altaffilmark{26},
 L.~A.~Antonelli\altaffilmark{27},
 P.~Antoranz\altaffilmark{28},
 M.~Backes\altaffilmark{29},
 J.~A.~Barrio\altaffilmark{30},
 D.~Bastieri\altaffilmark{31},
 J.~Becerra Gonz\'alez\altaffilmark{32,}\altaffilmark{33},
 W.~Bednarek\altaffilmark{34},
 A.~Berdyugin\altaffilmark{35},
 K.~Berger\altaffilmark{32}
% ...
\newline
(the MAGIC Collaboration)
\newline
D.~Paneque\altaffilmark{38,}\altaffilmark{51},
M.~Hayashida\altaffilmark{51}
}

\altaffiltext{1}{Fred Lawrence Whipple Observatory, Harvard-Smithsonian Center for Astrophysics, Amado, AZ 85645, USA}
\altaffiltext{2}{Department of Physics and Astronomy, University of California, Los Angeles, CA 90095, USA}
\altaffiltext{3}{Santa Cruz Institute for Particle Physics and Department of Physics, University of California, Santa Cruz, CA 95064, USA}
\altaffiltext{4}{Department of Physics, Washington University, St. Louis, MO 63130, USA}
\altaffiltext{5}{Astrophysical Institute, Department of Physics and Astronomy, Ohio University, Athens, OH 45701}
\altaffiltext{6}{Department of Physics and Astronomy and the Bartol Research Institute, University of Delaware, Newark, DE 19716, USA}
\altaffiltext{7}{School of Physics and Astronomy, University of Leeds, Leeds, LS2 9JT, UK}
\altaffiltext{8}{School of Physics, University College Dublin, Belfield, Dublin 4, Ireland}
\altaffiltext{9}{School of Physics, National University of Ireland Galway, University Road, Galway, Ireland}
\altaffiltext{10}{Astronomy Department, Adler Planetarium and Astronomy Museum, Chicago, IL 60605, USA}
\altaffiltext{11}{Department of Physics, Purdue University, West Lafayette, IN 47907, USA }
\altaffiltext{12}{Department of Physics, Grinnell College, Grinnell, IA 50112-1690, USA}
\altaffiltext{13}{Department of Physics and Astronomy, Barnard College, Columbia University, NY 10027, USA}
\altaffiltext{14}{Department of Astronomy and Astrophysics, 525 Davey Lab, Pennsylvania State University, University Park, PA 16802, USA}
\altaffiltext{15}{Department of Physics and Astronomy, University of Utah, Salt Lake City, UT 84112, USA}
\altaffiltext{16}{Physics Department, McGill University, Montreal, QC H3A 2T8, Canada}
\altaffiltext{17}{Department of Physics, Pittsburg State University, 1701 South Broadway, Pittsburg, KS 66762, USA}
\altaffiltext{18}{Enrico Fermi Institute, University of Chicago, Chicago, IL 60637, USA}
\altaffiltext{19}{Department of Physics and Astronomy, Iowa State University, Ames, IA 50011, USA}
\altaffiltext{20}{Department of Physics and Astronomy, University of Iowa, Van Allen Hall, Iowa City, IA 52242, USA}
\altaffiltext{21}{Department of Physics and Astronomy, DePauw University, Greencastle, IN 46135-0037, USA}
\altaffiltext{22}{Department of Life and Physical Sciences, Galway-Mayo Institute of Technology, Dublin Road, Galway, Ireland}
\altaffiltext{23}{Instituto de Astronomia y Fisica del Espacio, Casilla de Correo 67 - Sucursal 28, (C1428ZAA) Ciudad Aut—noma de Buenos Aires, Argentina}
\altaffiltext{24}{Kavli Institute for Cosmological Physics, University of Chicago, Chicago, IL 60637, USA}
\altaffiltext{25}{Department of Applied Physics and Instrumentation, Cork Institute of Technology, Bishopstown, Cork, Ireland}
\altaffiltext{26} {IFAE, Edifici Cn., Campus UAB, E-08193 Bellaterra, Spain}
 \altaffiltext{27} {INAF National Institute for Astrophysics, I-00136 Rome, Italy}
 \altaffiltext{28} {Universit\`a  di Siena, and INFN Pisa, I-53100 Siena, Italy}
 \altaffiltext{29} {Technische Universit\"at Dortmund, D-44221 Dortmund, Germany}
 \altaffiltext{30} {Universidad Complutense, E-28040 Madrid, Spain}
 \altaffiltext{31} {Universit\`a di Padova and INFN, I-35131 Padova, Italy}
 \altaffiltext{32} {Inst. de Astrof\'{\i}sica de Canarias, E-38200 La Laguna, Tenerife, Spain}
 \altaffiltext{33} {Depto. de Astrof\'{\i}sica, Universidad, E-38206 La Laguna, Tenerife, Spain}
 \altaffiltext{34} {University of \L\'od\'z, PL-90236 Lodz, Poland}
 \altaffiltext{35} {Tuorla Observatory, University of Turku, FI-21500 Piikki\"o, Finland}
% ...
\altaffiltext{38} {Max-Planck-Institut f\"ur Physik, D-80805 M\"unchen, Germany}
% ...
\altaffiltext{51}{W. W. Hansen Experimental Physics Laboratory, Kavli Institute for Particle Astrophysics and Cosmology, Department of Physics and SLAC National Accelerator Laboratory, Stanford University, Stanford, CA 94305, USA}
\altaffiltext{a}{Now at Department of Physics and Astronomy, University of Iowa, Van Allen Hall, Iowa City, IA 52242, USA}
\altaffiltext{b}{Now at DESY, Platanenallee 6, 15738 Zeuthen, Germany}
\altaffiltext{c}{Now at Institut f\"{u}r Physik und Astronomie, Universit\"{a}t Potsdam, 14476 Potsdam-Golm,Germany; DESY, Platanenallee 6, 15738 Zeuthen, Germany}
\altaffiltext{d}{Now at Los Alamos National Laboratory, MS H803, Los Alamos, NM 87545}
% ...
\altaffiltext{$\dagger$}{corresponding authors: Daniel Gall daniel-d-gall@uiowa.edu and Wei Cui cui@purdue.edu}

\begin{document}
%\linenumbers

\begin{abstract}
\noindent
The very high energy (VHE; $E$ $>$ 100 GeV) blazar
Markarian 501 has a well-studied history of extreme spectral variability and is an excellent laboratory for studying the physical processes within the jets of active galactic nuclei.  However, there are few detailed multiwavelength studies of Markarian 501 during its quiescent state, due to its low luminosity.  A short-term multiwavelength study of Markarian 501 was coordinated in March 2009, focusing around a multi-day observation with the Suzaku X-ray satellite and including $\gamma$-ray data from VERITAS, MAGIC, and the {\em Fermi} Gamma-ray Space Telescope with the goal of providing a well-sampled multiwavelength baseline measurement of Markarian 501 in the quiescent state.  The results of these quiescent-state observations are compared to the historically extreme outburst of April 16, 1997, with the goal of examining variability of the spectral energy distribution between the two states.  The derived broadband spectral energy distribution shows the characteristic double-peaked profile.  We find that the X-ray peak shifts by over two orders of magnitude in photon energy between the two flux states while the VHE peak varies little.  The limited shift in the VHE peak can be explained by the transition to the Klein-Nishina regime.  Synchrotron self-Compton models are matched to the data and the implied Klein-Nishina effects are explored.

\end{abstract}

\section{Introduction}

Blazars, a subclass of active galactic nuclei (AGN), are the dominant extragalactic source class in $\gamma$-rays. They have been observed to show rapid variability and non-thermal spectra, presenting a broad continuum across nearly the entire electromagnetic spectrum.  This implies that the observed photons originate within highly relativistic jets oriented very close to the observer's line of sight \citep{Urry}. This orientation results in Doppler beaming that boosts the intensity and frequency of the observed jet emission, often overwhelming all other emission from the source.  Therefore, blazars make excellent laboratories for studying the physical processes within the jets of AGN. They were among the first sources to be detected in the VHE band, and as of this writing there are 34 known VHE $\gamma$-ray blazars\footnote{http://tevcat.uchicago.edu/}.   Markarian 501 (Mrk 501; 1H1652+398), at a redshift of $z=0.034$, was the second blazar to be detected at VHE \citep{501discovery}.

The spectral energy distribution (SED) of Mrk 501 characteristically shows a double-peaked profile.  These peaks occur at keV and TeV energies when the SED is plotted in the $\nu$F$_\nu$ vs $\nu$ representation.  This structure is common among all VHE $\gamma$-ray blazars, and several models have been developed to account for the double-peaked structure.  These models uniformly attribute the peak at keV energies to synchrotron radiation from relativistic electrons and positrons within the blazar jets, but they differ in accounting for the source of the VHE peak.  The models are generally divided into two classes:  leptonic and hadronic, named for their attributed source for the VHE peak.  The leptonic models advocate inverse-Compton scattering to VHE of either the synchrotron photons from within the jet or an external photon field \citep[e.g.,][]{Marscher,Maraschi,Dermer,Sikora}.  The hadronic models, however, account for the VHE emission by $\pi^0$ decay or by $\pi^{\pm}$ decay with subsequent synchrotron and/or Compton emission from decay products, or by synchrotron radiation from ultra-relativistic protons \citep[e.g.,][]{Mannheim, Aharonian, Pohl}.

Observationally, Mrk 501 has been known to undergo both major outbursts on long time scales and rapid flares on short time scales, most prominently in the keV and VHE range \citep[e.g.,][]{Catanese, Pian, xue501, albert}. During these outbursts, both of the SED peaks have been observed to shift towards higher energies.  During the most extreme cases, the keV peak has been observed above 200 keV, well above typical values below 1 keV.  Historically, the SED has been measured in the VHE band primarily during outbursts, due to the lower sensitivity of previous generations of instruments.  A previous study examined the quiet state, but it was performed before the launch of the Fermi Large Area Telescope (\FermiLATc), which provides coverage in the range between keV and VHE energies \citep{magic501}.  This work attempts to provide state-of-the-art short-term multiwavelength measurements of the quiescent state of Mrk 501, with broad spectral coverage in the critical keV and VHE bands as well as coverage with the \FermiLATc.  These measurements are then compared to observations by BeppoSAX in X-rays and by the Whipple 10m and CAT Cerenkov telescopes in VHE $\gamma$-rays during the 1997 extreme outburst.  This outburst has been well studied using multiple instruments \citep[e.g.,][]{hegra501, Catanese, hegra501, Pian} and provides a good comparison to the quiet state observed in 2009.

\section{Data Selection and Analysis}
\subsection{Suzaku:  X-ray}
The Suzaku X-ray observatory, a collaborative project between institutions in the United States and Japan, is an excellent tool for studying the broadband SED of sources such as Mrk 501 due to its broad energy range (0.2 - 600 keV).  Suzaku has two operating instruments for studying X-ray emission: the X-ray imaging spectrometer \citep[XIS;][]{xis} and the hard X-ray detector \citep[HXD;][]{hxd}. 

The XIS instrument consists of four X-ray telescopes with imaging CCD cameras at their focal planes.  Three of the CCDs are front-illuminated, and one CCD is back-illuminated to provide extended sensitivity to soft X-rays.  The combined energy range of these CCDs is 0.2 - 12.0 keV.  The HXD is a non-imaging instrument that expands the energy sensitivity of Suzaku to the 10 - 600 keV band by using two types of detectors.  Silicon PIN diodes provide sensitivity in the range 10 - 70 keV, and gadolinium silicate [GSO; Gd$_2$SiO$_5$(Ce)] scintillators placed behind the PIN diodes provide sensitivity in the range 40 - 600 keV.  One limitation of Suzaku with respect to attaining simultaneous multiwavelength observations is that it resides in a low earth orbit, so observations are subject to frequent earth occultations, limiting continuous temporal coverage.

Observations of Mrk 501 were carried out with the Suzaku X-ray satellite from 2009-03-23 UT 18:39 to 2009-03-25 UT 07:59 (sequence number 703046010).  After good time interval selection and dead-time correction, a total of approximately 72 ks of live time remained for the XIS and 61 ks for the HXD.  To reduce the occurrence of photon pile-up in the XIS, a common problem when dealing with bright X-ray point sources, the XIS observations were carried out in 1/4 window mode and Suzaku was operated with a pointing nominally centered on the HXD field of view (FoV).  In addition, HXD nominal pointing increases the effective area of the HXD instrument.  The resulting XIS count rates are well below the threshold where pile-up becomes a significant issue.  Even with the HXD nominal pointing, there was no significant detection in the GSO scintillators, so the maximum useful energy for these observations using only the HXD/PIN detector was around 70 keV.  

The raw XIS data were reprocessed following the guidelines from the Suzaku team\footnote{http://heasarc.gsfc.nasa.gov/docs/suzaku/analysis/abc/node9.html} using Suzaku FTOOLS version 12.0.  The events were graded and only those with ASCA grades 0, 2, 3, 4 and 6 were included in further analysis.  In addition, the data were screened to exclude times when the Suzaku spacecraft was passing through the South Atlantic Anomaly (SAA), including a 436 s post-SAA buffer to allow the instrument backgrounds to settle.  Additional cuts were made requiring the data used to be taken at least $5{\degree}$ in elevation above the Earth's limb and at least $20{\degree}$ above the Earth's daytime limb.  These cuts help reduce contamination from the Earth's atmosphere.  

Source events were extracted from a circular region with a radius of 1.92' centered on Mrk 501.  Background events were extracted from a circular region with the same radius located far from the source in each XIS detector.  The redistribution matrix file (RMF) and ancillary response files (ARF) were generated for each XIS detector using the analysis tools included in the Suzaku analysis package.

The HXD data were also reprocessed from the raw files, following the guidelines provided by the Suzaku team.\footnote{http://heasarc.gsfc.nasa.gov/docs/suzaku/analysis/hxd\_repro.html}  The data were filtered with similar criteria to the XIS data with a slightly longer settling time of 500 s required following passage through the SAA.  In addition, the only elevation requirement was for data to be taken at least $5{\degree}$ above the Earth's limb.

Because the HXD is a non-imaging instrument, the cosmic X-ray background (CXB) and instrumental non-X-ray background (NXB) must be accounted for in different ways than for the XIS.  The time-variable non-X-ray background has been modeled and provided by the HXD team.  The cosmic X-ray background is not included in this model and must be accounted for separately as the CXB flux is about 5$\%$ of the background for PIN observations.  The model suggested by the HXD team is based on results from HEAO-1 and modified for input to XSPEC.  The imported model reproduces the PIN background with an HXD nominal pointing \citep{boldt}.  The model is given by:

\begin{equation} \frac{dF}{dE}=8.0\times10^{-4} \left( \frac{E}{1\mbox{ }\mathrm{keV}}\right)^{-1.29}\times \exp \left( \frac {-E}{40\mbox{ }\mathrm{keV}} \right) \mathrm{photons}\mbox{ } \mathrm{cm}^{-2}\mbox{ }  \mathrm{s}^{-1}\mbox{ }  \mathrm{FOV}^{-1}\mbox{ } \mathrm{keV}^{-1}\;.
\label{eqn:pinbgd}
 \end{equation}

The CXB background model was added to the NXB background model and used to subtract background events in XSPEC during spectral analysis of the PIN data.  The response file \texttt{ae\_hxd\_pinhxnome5\_20080716.rsp} provided by the HXD team was used for the PIN spectral analysis.

\subsection{VERITAS:  VHE $\gamma$-ray}

The Very Energetic Radiation Imaging Telescope Array System (VERITAS) is an array of four imaging atmospheric-Cerenkov telescopes (IACTs) located at the Fred Lawrence Whipple Observatory in southern Arizona (31$^{\circ}$40'N, 110$^{\circ}$57'W) at an altitude of 1.28 km above sea level \citep{veritasstatus}.  These telescopes capture the Cerenkov light emitted by extensive air showers that are initiated in the upper atmosphere by $\gamma$-rays and cosmic rays.  The telescopes have 12m-diameter tessellated reflectors directing light onto imaging cameras, each consisting of 499 photomultiplier tubes (PMTs), located in the focal plane of each telescope.  The cameras have a FoV of 3.5$^{\circ}$ with an angular size per pixel of 0.15$^{\circ}$.  VERITAS is sensitive in the range of approximately 100 GeV -- 30 TeV with an energy resolution of between 15 and 20\%.  Images of the extensive air showers captured by each telescope can be combined to reconstruct the direction and energy of the shower initiator.  The telescopes typically operate in ``wobble" mode, where the location of the object to be observed is offset from the center of the FoV by $0.5^{\circ}$, allowing for simultaneous source and background measurements \citep{Fomin}.  The offset direction alternates between north, south, east and west for each data run (typically lasting 20 minutes) to reduce systematic effects from the offset.

The VERITAS data were analyzed using the standard analysis package \citep{vegascogan}.  Images of the Cerenkov light produced by the extensive air showers are parameterized, and quality cuts are placed on the data to ensure that the shower reconstruction procedure can be performed reliably.  Separation of showers initiated by $\gamma$-ray photons from hadronic showers initiated by cosmic rays is performed with the aid of lookup tables.  The lookup tables are calculated for each telescope with several values of zenith angle, azimuth angle, and the night sky background noise level.  The observed parameters \emph{length} and \emph{width} are scaled by the corresponding parameters in the lookup tables and averaged among the telescopes to find the ``mean scaled parameters" in the method first suggested by \citet{daumwobble}.  Selection criteria for $\gamma$/hadron separation (optimized with data taken on the Crab Nebula) are placed on the mean scaled parameters of the shower images to exclude most hadronic showers while still retaining a large proportion of $\gamma$-ray initiated showers.

To perform a background subtraction of the surviving cosmic-ray events, an estimation of these background counts is made using the ``reflected-region" background model \citep{Berge}.  Events within a squared angular distance of $\theta^{2}<0.0169$ of the anticipated source location are considered ON events.  Background measurements (OFF events) are taken from regions of the same size and at the same angular distance from the center of the FoV.  For this analysis, a minimum of eight background regions were used.  The excess number of events from the anticipated source location is found by subtracting the number of OFF events (scaled by the relative exposure) from the ON events.  Statistical significances are calculated using Eqn. 17 of \citet{LiMa}.  More details about VERITAS, the calibration and analysis techniques can be found in \citet{veritas}.

On March 24, 2009 (MJD 54914), VERITAS was used to take data on Mrk 501 while operating in wobble mode, with an offset angle of 0.5$\degree$.  VERITAS was used to observe Mrk 501 for 2.6 hours from 9:11 UT to 11:58 UT using a series of 20-minute runs, with a live-time of 2.4 hours and zenith angles ranging from 35$^{\circ}$ to 9$^{\circ}$.  Additional observations totaling 2.9 hours were made on March 23 and 25 but were discarded due to poor weather conditions.  The data were processed using standard VERITAS analysis cuts, finding 431 ON events and 1810 OFF events with a maximum ratio between ON and OFF exposure of 0.125.  The total significance of the excess for this observation was 12.7 standard deviations ($\sigma$) with an average significance per run of 3.8 $\sigma$.

\subsection{MAGIC:  VHE $\gamma$-ray}

MAGIC is a system of two IACTs for VHE $\gamma$-ray astronomy located on the Canary island of La Palma (2.2 km above sea level, 28$^{\circ}$45'N, 17$^{\circ}$54'W).
At the time of the observation, MAGIC-II, the second telescope of the array, was still in its commissioning phase so Mrk501 was observed by MAGIC-I in stand-alone mode.

MAGIC-I has a 17m-diameter tessellated reflector dish and a hexagonal camera with a FoV of 3.5$^{\circ}$ mean angular diameter.
The camera comprises 576 high-sensitivity PMTs of different diameters (0.1$^{\circ}$ in the inner portion of the camera and 0.2$^{\circ}$ in the outer portion).
The trigger threshold of MAGIC at low zenith angles is between 50 and 60 GeV with the standard trigger configuration used in this campaign. 
The accessible energy range extends up to tens of TeV with a typical energy resolution of 20\% to 30\%, depending on zenith angle and energy.
Further details, telescope parameters, and performance information can be found in \citet{albertA}.

The MAGIC data for this campaign were recorded between 3:50 UT and 5:10 UT on the night of March 23, 2009 (MJD 54913). 
The observations were performed in wobble mode. 
The zenith angle of the observations ranged from 15$^{\circ}$ to 30$^{\circ}$. 
The recorded data fulfil the standard quality requirements, so none of the runs were removed from the data sample. The total live time for this observation is 1.3 h.

The data were analyzed using the standard MAGIC calibration and analysis \citep{albertA}. 
The analysis is based on image parameters \citep[]{hillas,Aliu09} and the random forest (RF) method \citep{albertB}, which are used to define the so-called hadronness of each event. 
The cut in hadronness for background rejection was chosen using Monte Carlo data, setting a cut efficiency for $\gamma$-rays of 70\%.
Three background (OFF) sky regions are used for the residual background estimation, giving a ratio between ON and OFF exposure of 0.333.
The OFF regions are chosen to be at the same angular distance from the camera center as the ON region. 
The excess of $\gamma$-ray events from the source was extracted from the distribution of the ALPHA parameter, which is related to the image orientation on the camera \citep{hillas}.
The final cut on ALPHA was also chosen using Monte Carlo $\gamma$-ray events, again imposing a cut efficiency of 70\%.
With these less restrictive cuts, suitable for the spectrum calculation, the significance of the excess is 2.2 $\sigma$ with 9527 ON events and 27846 OFF events, whereas applying stricter cuts in the optimal energy range for signal detection leads to a significance of 9.7 $\sigma$ with 123 ON events and 97 OFF events.

The energies of the VHE $\gamma$-ray candidates were also estimated using the RF method. 
A cut was applied on the SIZE parameter, requiring a minimum of 100 photoelectrons in the shower images.  Because the SIZE parameter is related to the energy of the initiator of the shower, this cut results in an energy threshold for the analysis (defined as the peak position of the distribution of the reconstructed energy parameter for simulated $\gamma$-ray events after the cut) of $\sim$ 100 GeV.

\subsection{{\em Fermi}-LAT:  HE $\gamma$-ray}

The {\em Fermi} Large Area Telescope (LAT) is a satellite instrument to perform $\gamma$-ray astronomy from $20$\,MeV to several hundred GeV. The instrument is an array of $4 \times 4$ identical towers, each one consisting of a tracker (where the photons are pair-converted) and a calorimeter (where the energies of the pair-converted photons are measured). The entire instrument is covered with an anticoincidence detector to reject charged-particle background. The LAT has a peak effective area of $0.8$\,m$^2$ for $1$\,GeV photons, energy resolution typically better than $10\%$, and a FoV of about $2.4$\,sr, with angular resolution  ($68\%$ containment angle) better than $1^{\circ}$ for energies above $1$\,GeV. Further details about the LAT can be found in \cite{Atwood2009}

The analysis was performed with the ScienceTools software package version v9r15p2,  which is available from Fermi Science Support Center (FSSC). Only events having the highest probabilities of being photons, those in the ``diffuse'' class,  were used. The Pass 6v3 diffuse-class instrument response functions were used for the analysis.  The LAT data were extracted from a circular region of $10^{\circ}$ radius centered at the location of Mrk501.
The spectral fit was performed using photon energies greater than $0.3$ GeV in order to minimize systematic effects, and a cut on the zenith angle ($> 105^{\circ}$) was also applied to reduce contamination from Earth-albedo $\gamma$-rays, which are produced by cosmic rays interacting with the upper atmosphere.

The background model used to extract the $\gamma$-ray signal includes
a Galactic diffuse emission component and an isotropic
component.  The model that we adopted for the Galactic component is  gll\_iem\_v02.fit \footnote{
  http://fermi.gsfc.nasa.gov/ssc/data/access/lat/BackgroundModels.html}.
The isotropic component, which is the sum of the extragalactic
diffuse emission and the residual charged-particle background, is
parametrized here with a single power-law function.  Owing to the relatively small size of the region analyzed (radius 10$^\circ$) and the hardness of the spectrum of Mrk501, the high-energy structure in the standard tabulated isotropic background spectrum, isotropic\_iem\_v02.txt, does not dominate the total counts at high energies.  In addition, we find that for this region a power-law approximation to the isotropic background results in somewhat smaller residuals for the model overall, possibly because the isotropic term, with a free spectral index, compensates for an inaccuracy in the model for the Galactic diffuse emission, which is also approximately isotropic at the high Galactic latitude of Mrk501 ($b \sim 39^\circ$).  In any case, the resulting spectral fits for Mrk501 are not significantly different if isotropic\_iem\_v02.txt is used for the analysis.  To further reduce systematic uncertainties in the
analysis, the photon index of the isotropic component and the normalization of both components in the background
model were allowed to vary freely during the spectral point fitting.
In addition, the model also includes five nearby sources from the 1FGL catalog \citep{1FGL}: 1FGL J1724.0+4002, 1FGL J1642.5+3947, 1FGL J1635.0+3808, 1FGL J1734.4+3859, 1FGL J1709.6+4320. The spectra of those sources were also parameterized by power-law functions, whose photon index values were fixed to the values from the 1FGL catalogue, and only the normalization factors for the single sources were left as free parameters. 
The spectral analysis was performed with the post-launch instrument-response functions \texttt{P6\_V3\_DIFFUSE} using an unbinned maximum
likelihood method \citep[]{Mattox1996}. The \Fermi collaboration estimates the current
systematic uncertainties on the flux as $10\%$ at $0.1$\,GeV, $5\%$ at $560$\,MeV and $20\%$ at $10$\,GeV and above\footnote{See \texttt{http://fermi.gsfc.nasa.gov/ssc/data/analysis/LAT\us caveats.html}}.

\section{Results}

\subsection{Light Curves}
\label{sec:lightcurves}
\begin{figure}[htbp]
\begin{center}
\includegraphics[width=4in]{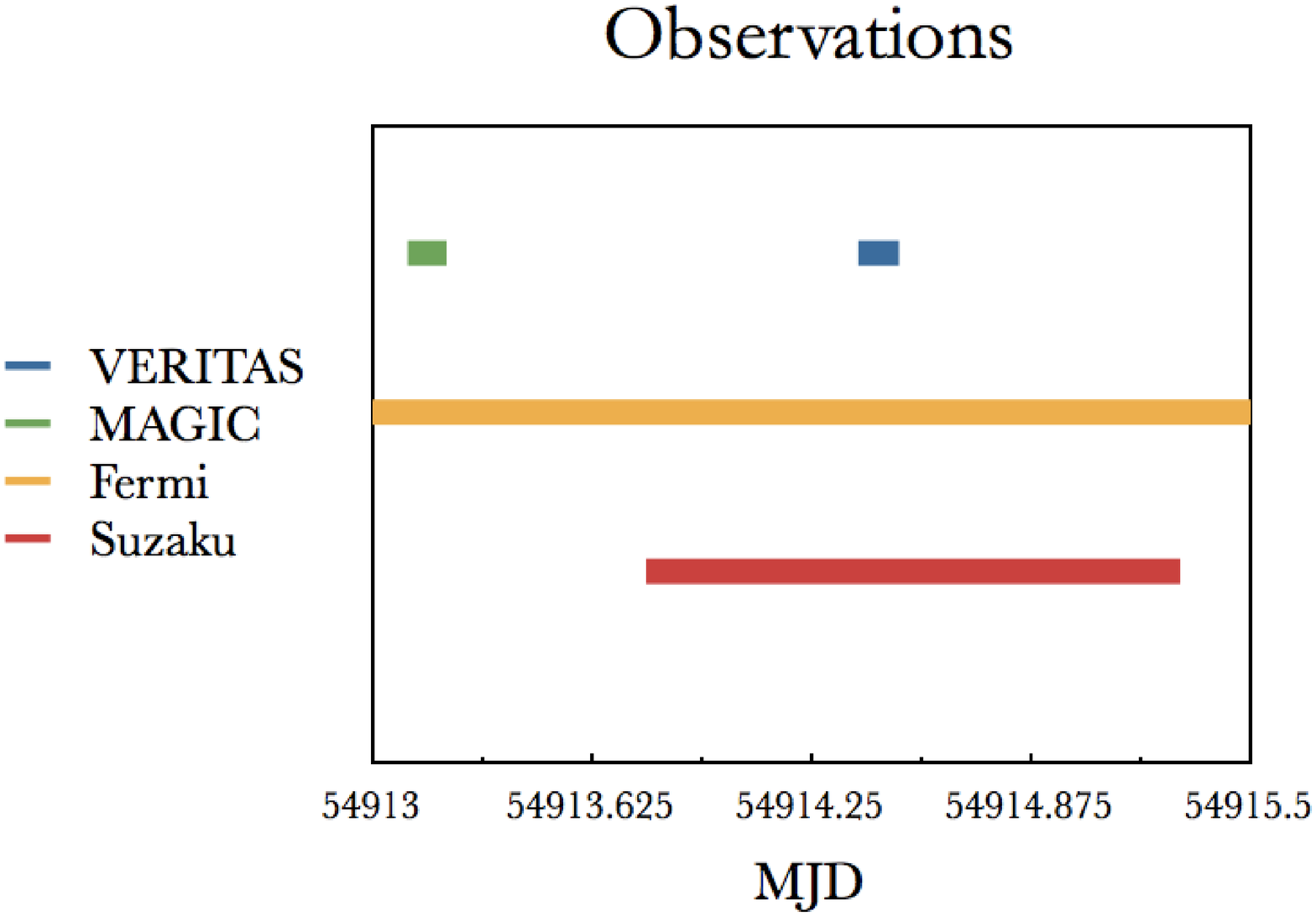}
\caption{{Overall time coverage of Mrk 501 by the instruments involved in the multiwavelength campaign.  Gaps due to orbital constraints are not indicated in the Fermi and Suzaku plots.  A color version of this figure is available online.}}
\label{fig:observations}
\end{center}
\end{figure}

\begin{figure}[t]
\centering
\includegraphics[width=80mm]{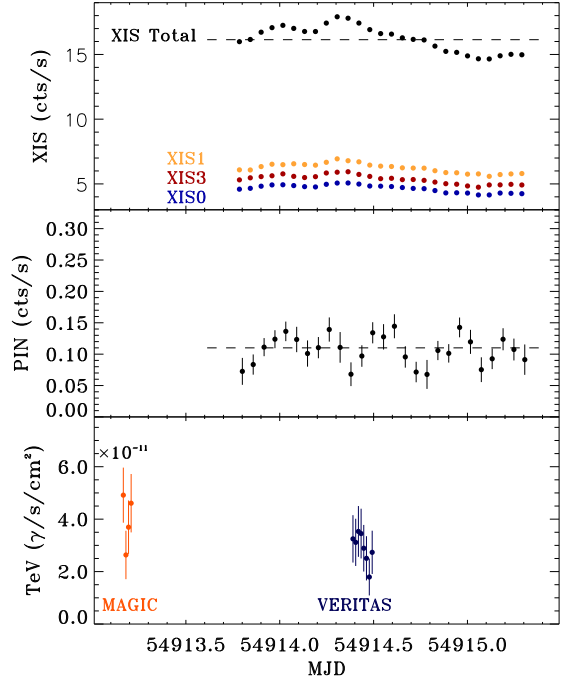}
\caption{Light curves of Mrk 501 in the energy ranges used for Suzaku/XIS (0.6 keV $<$ E $<$ 10.0 keV), Suzaku/HXD (10 keV $<$ E $<$ 70 keV), MAGIC and VERITAS (E $>$ 300 GeV). Error bars show 1$\sigma$ statistical errors.  The constant fits to the Suzaku data are shown.  See discussion in Sec. \ref{sec:lightcurves}. A color version of this figure is available online.} \label{lc}
\end{figure}

The time coverage of the data sets used for this campaign is illustrated in Figure \ref{fig:observations}.  Light curves were produced in the passband of each instrument involved in the campaign to understand the flux levels and variability of Mrk501 during these observations; however, the \FermiLATc \mbox{ }detection was not significant enough to include a light curve.  The resulting multiwavelength light curves are shown in Figure \ref{lc}.  

Light curves produced from the XIS and HXD/PIN Suzaku data, with 5 ks time bins, are shown in the top two panels of Figure \ref{lc}.  Individual light curves for each of the XIS detectors were created and then summed.  The best fit of a constant for the summed XIS (0.6 -- 10.0 keV) light curve is 16.14 counts s$^{-1}$ with a $\chi^{2}$ value of 4207 for 26 degrees of freedom.  The poor fit of a constant to the data corresponds to the moderate variability that is visible by inspection in the XIS count rate.  A constant value was also fitted to the HXD/PIN light curve, yielding a value of 0.11 counts s$^{-1}$ with a $\chi^{2}$ value of 43.7 for 26 degrees of freedom, again indicating variability.

In the VHE $\gamma$-ray band, the source appears to be in a ``quiescent" state when compared to contemporaneous long-term studies of the source's broadband behavior \citep{paneque, Huang, Pichel}.  The light curves produced from the VERITAS and MAGIC data used 20 minute time bins and are plotted in the bottom panel of Figure \ref{lc}.  To produce light curves, the spectral slope from the corresponding data is assumed (see Section \ref{sec:spectrum}), and the effective collection areas were used with the excess counts to determine the spectral normalization.  An energy threshold of 300 GeV was then used to calculate the integral flux for each bin.  A constant flux value was fitted to the VERITAS and MAGIC data separately, yielding a best fit value of $(2.8\pm0.5)\times10^{-11}\mbox{ } \mathrm{photons} \mbox{ } \mathrm{cm}^{-2} \mbox{ } \mathrm{s}^{-1}$  for VERITAS and $(3.9\pm0.8)\times10^{-11}\mbox{ } \mathrm{photons} \mbox{ } \mathrm{cm}^{-2} \mbox{ } \mathrm{s}^{-1}$ for MAGIC.  The VERITAS data show no evidence for any variability, with a $\chi^{2}$ value of 3.6 for 7 degrees of freedom for the constant fit.  Likewise, the constant fit to the MAGIC data yields a $\chi^{2}$ value of 3.2 for 3 degrees of freedom.  The individual constant fits to the VERITAS and MAGIC data differ slightly but show that the flux measurements from the two instruments are reasonably compatible, with no large-scale variability.

\subsection{Spectral Analysis and Modeling}
\label{sec:spectrum}
\begin{figure}[t]
\centering
\includegraphics[width=80mm, angle=-90]{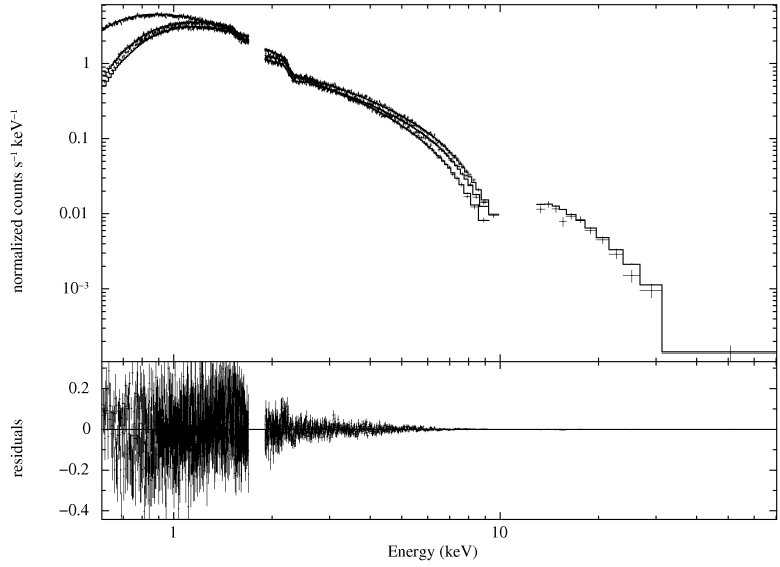}
\caption{Observed X-ray spectra of Mrk 501 made from Suzaku XIS and HXD/PIN data separately.  A joint fit to a broken power-law model was performed for the three XIS CCDs and the HXD/PIN.  The solid histogram shows the best fits to the data (see Sec. \ref{sec:spectrum}).  The error bars represent 1$\sigma$ statistical errors.  The bottom panel shows the residuals of the fit. } 
\label{fig:suzakuspec}
\end{figure}

\begin{figure}[t]
\centering
\includegraphics[width=80mm]{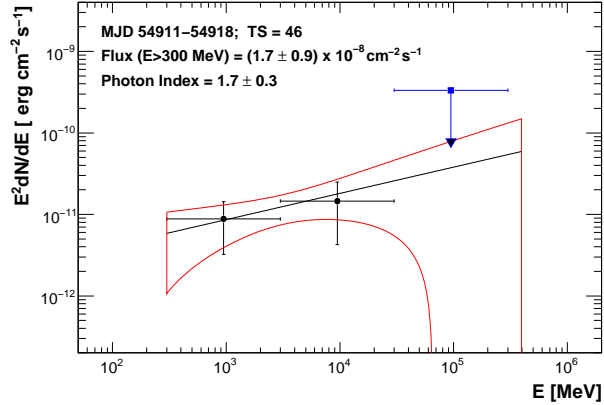}
\caption{Plot of the Fermi spectral results for the 7 days centered on the Suzaku observing time window (MJD 54911 -- MJD 54918).  The solid line depicts the result of the unbinned likelihood power-law fit, and the contour shows the 68\% uncertainty of the fit.  The data points show the energy fluxes computed in differential energy ranges, with a 95\% confidence level upper limit for the highest energy range.  The Fermi analysis was performed in the energy range 0.3-400 GeV.  The vertical bars show 1$\sigma$ statistical errors, and the horizontal bars show the energy range for each point.  A color version of this figure is available online.}
\label{fig:fermispec}
\end{figure}

\begin{figure}[t]
\centering
\includegraphics[width=80mm]{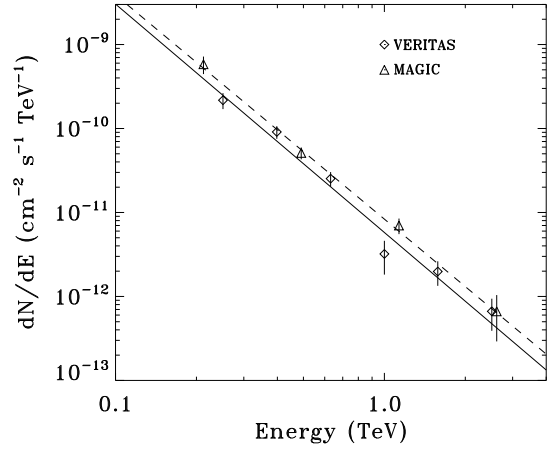}
\caption{VHE spectrum from VERITAS (diamonds) and MAGIC (triangles), including the best-fit power-law models.  The VERITAS fit is shown with a solid line, and the MAGIC fit is shown with a dashed line.  The slopes of the power-law models agree well ($\Gamma_{VERITAS}=2.72\pm0.15_{stat}\pm0.10_{sys}$ and $\Gamma_{MAGIC}=  2.67\pm0.21_{stat}\pm0.20_{sys}$).  The slight normalization offset corresponds to the moderately higher VHE flux levels during the MAGIC observation.  The vertical bars show 1$\sigma$ statistical errors.} \label{tevspec}
\end{figure}

\begin{figure}[t]
\centering
\includegraphics[width=160mm]{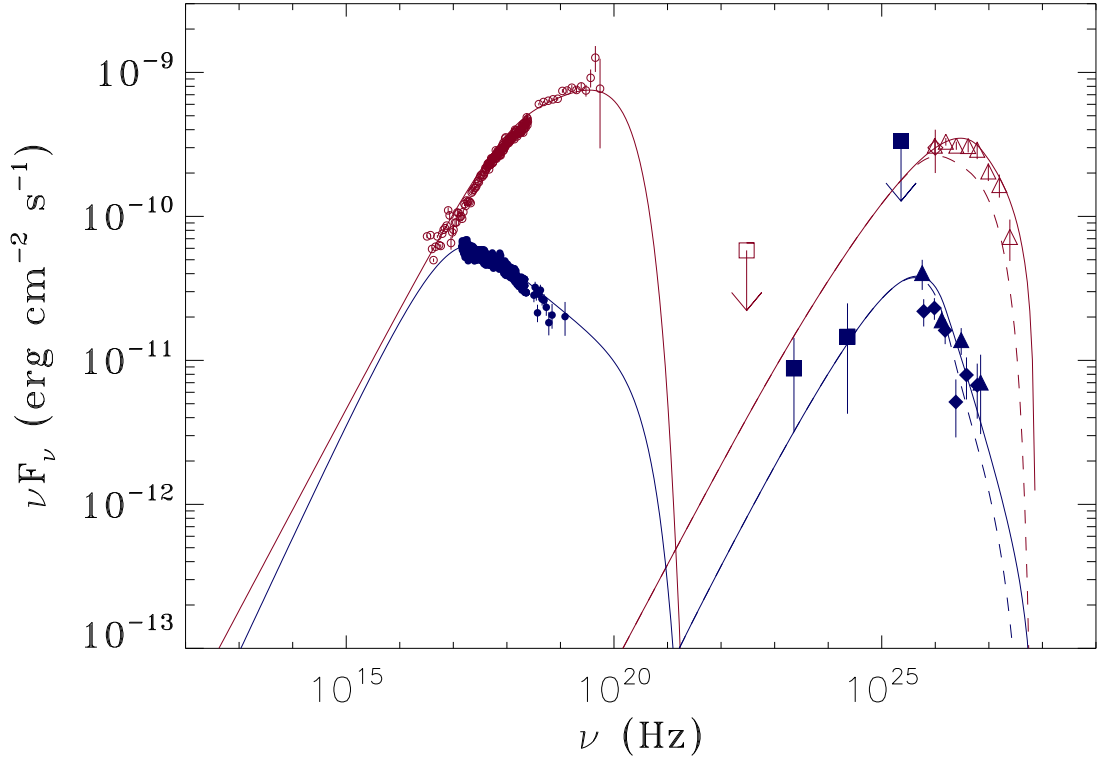}
\caption{Spectral energy distributions of Mrk 501 in the quiescent state (in filled points) and during the 1997 extreme outburst (in open points).  The former is made from data obtained in the present campaign with Suzaku (filled circles), Fermi (filled squares), VERITAS (filled diamonds), and MAGIC (filled triangles).  The latter was taken from \citet{Pian} and \cite{cat}, with X-ray data from BeppoSAX (open circles), GeV $\gamma$-ray upper limit from EGRET (open square), and VHE $\gamma$-ray data from the Whipple 10m (open diamond) and the CAT Cerenkov telescope (open triangles).  Note that the Whipple 10m spectral point overlaps with the lowest-energy CAT spectral point.  The vertical bars for all data show 1$\sigma$ statistical errors and the upper limits are at the 95\% confidence level.  Corresponding SED models matched to the data for each year are shown.  The solid lines show the SSC match; dashed lines take EBL absorption into account.  The VHE data points are not EBL-corrected.  A color version of this figure is available online.} 
\label{fig:sed}
\end{figure}

The spectral results for the observations taken during this campaign when Mrk 501 was in the quiescent  or low-flux state are shown in Figures \ref{fig:suzakuspec}, \ref{fig:fermispec} and  \ref{tevspec}.  For comparison, they are also shown in Figure \ref{fig:sed} along with archival data from \cite{Pian}, \cite{Catanese}, and \cite{cat} taken when Mrk 501 was in a high-flux state.

The data from all Suzaku instruments were fitted jointly in XSPEC 12 \citep{Arnaud}.  A cross-normalization parameter was introduced to account for the calibration differences between the individual XIS detectors and the HXD/PIN.  A broken power-law function modified by interstellar absorption with a fixed column density N$_H=1.73\times10^{20}$ cm$^{-2}$  \citep{stark} was fitted to the data, yielding power-law indices $\Gamma_1=2.133\pm0.003_{stat}$ and $\Gamma_2=2.375\pm0.009_{stat}$, and a break energy of $E_{break}=3.21\pm0.07_{stat}$ keV.  After the fitting procedure, the cross-normalization factors for the XIS1 and XIS3 detectors relative to the XIS0 detector are compatible with previous calibration measurements of the Crab Nebula flux \citep{xrt}.  A fixed normalization factor of 1.18 was introduced between the HXD/PIN and XIS0 as specified by the Suzaku team for HXD nominal pointings \citep{maeda}.  The resulting fit, with $\chi^2$/d.o.f.$=1.33$ for 1511 degrees of freedom, is plotted in Figure \ref{fig:suzakuspec}.  After an acceptable fit was found, the result was unfolded through the instrument response and de-absorbed to derive the intrinsic X-ray spectrum shown in Figure \ref{fig:sed}.

The \FermiLAT spectral measurement shown in Figures \ref{fig:fermispec} and \ref{fig:sed} is derived using a time interval  of 7 days centered on the Suzaku observing time window -- from MJD 54911 up to MJD 54918. Mrk501 is detected with a Test Statistic (TS) value of 46 ($\sim7 \sigma$). The spectrum can be fitted with a single power-law function with photon flux  $F(>0.3 \;\mathrm{GeV})=(1.7 \pm 0.9_{stat}) \times 10^{-8}$ ph cm$^{-2}$s$^{-1}$ and differential photon spectral index   $\Gamma_{LAT}=1.7 \pm 0.3_{stat}$.  The number of photons (predicted by the model) for the entire spectrum is 12, out of which 9 are in the energy range 0.3-3 GeV, 2 in the energy range 3-30 GeV and 1 in the energy range 30-300 GeV.  In order to increase the simultaneity of the reported SED, we also performed the analysis of the LAT data over a period of 3 days centered at the Suzaku observations, namely the time interval from  MJD 54913 to MJD 54916. For such a short time interval the detection has a TS value of 10 ($\sim$ 3$\sigma$), and significantly greater uncertainties in the power-law parameters: $F(>0.3  \;\mathrm{GeV}) = (1.8 \pm 1.3_{stat}) \times 10^{-8}$ph cm$^{-2}$s$^{-1}$, $\Gamma_{LAT}=1.9 \pm 0.5_{stat}$. Since the two spectral shapes are compatible within errors and there is no indication of flux variability during this time, the spectrum derived with the 7-day time interval is used in the SED plot in Figures \ref{fig:fermispec} and \ref{fig:sed}.

For the VERITAS spectral analysis, the spectrum was unfolded using effective collection areas for the array that were calculated using \emph{Monte Carlo} simulations of extensive air showers passed through the analysis chain with a configuration corresponding to the data-taking conditions and the $\gamma$/hadron selection cuts.  The unfolding process takes into account the energy resolution of the array and potential energy bias.  The effective areas are then used for each reconstructed event to calculate the differential photon flux.  The VHE $\gamma$-ray differential spectral points were fit with a simple power law of the form:
\begin{equation}
\frac{dN}{dE} = F_0 \times 10^{-12} \times
 \left(\frac{E}{1\mbox{ }TeV}\right)^{-\Gamma_{VHE}}
 \mathrm{photons\mbox{ } cm}^{-2}\mbox{ } \mathrm{s}^{-1} \mbox{ }\mathrm{TeV}^{-1}\;.
 \label{eqn:tevspec}
\end{equation}

\noindent For the VERITAS analysis, best-fit parameters of $\Gamma_{VHE} = 2.72\pm0.15_{stat}\pm0.1_{sys}$ and $F_0=5.78\pm0.83_{stat}\pm1.16_{sys}$ were found.  

For the MAGIC analysis, the derived spectrum was unfolded to correct for the effects of the energy resolution of the detector \citep{albert2007} and of possible bias.  Four statistically significant spectral bins were obtained.  The spectrum is compatible with a simple power-law function with $\Gamma_{VHE}= 2.67 \pm 0.21_{stat} \pm 0.20_{sys}$ and $F_0=8.34 \pm1.53_{stat} \pm 2.50_{sys}$ within the MAGIC energy range (see Figure \ref{tevspec}).

A simple synchrotron self-Compton (SSC) model\footnote{The SSC code is available at http://jelley.wustl.edu/multiwave/spectrum/?code} is used to match the results with an emission scenario \citep{ssccode}.  This model assumes a spherical emission region of radius $R$ with the emission Doppler-boosted by the factor:
\begin{equation}
D=[\Gamma_{Lorentz}(1-\beta cos\theta)]^{-1}\;,
 \label{eqn:doppler}
\end{equation}

\noindent where the emission region is moving with Lorentz factor $\Gamma_{Lorentz}$ towards the observer, $\beta$ gives the bulk velocity of the plasma in terms of the speed of light, and $\theta$ is the angle between the jet axis and the line of sight in the observer frame.  The emission region contains an isotropic non-thermal electron population and randomly oriented magnetic field, $B$.  The electron spectrum is assumed to be in the form of a broken power-law function with index $p_1$ between $\gamma_{min}$ and $\gamma_{b}$ and index $p_2$ between $\gamma_{b}$ and $\gamma_{max}$.  The model accounts for $\gamma$-ray absorption by the extragalactic background light \citep{EBLfranceschini}.  After computing the synchrotron and inverse-Compton emission, the model corrects the photon spectrum that is produced for synchrotron-self-absorption and internal absorption due to pair-production processes. The model does not evolve the electron spectrum self-consistently.

In general, it is difficult to fully constrain the parameters in SSC models.  This requires both the flux and energies of the two (synchrotron and SSC) peaks to be fully determined by observations as well as an independent estimate of the Doppler factor \citep{ghisellini}.  In addition, measurements of the synchrotron spectrum are used to help determine the indices of the modeled electron spectrum ($p_1$ and $p_2$).  The SEDs from both sets of observations presented here constrain the synchrotron and SSC peak locations well (particularly for the low-flux state).  In addition, $p_1$ for the 1997 high-flux state and $p_2$ for the 2009 low flux-state are well constrained by the data, while the other electron spectral indices are not as restricted.  The parameters for the models used in this work for the 1997 high-flux state and the 2009 low-flux state are given in Table \ref{tab:sscmodels} along with model parameters from other works for comparison.  Some of the model parameters were matched to previous models as discussed below.

\section{Discussion}
This data set provides a high-quality sampling of the broadband SED of Mrk 501 in the quiescent state, and it allows comparisons to be made over a broad energy range with the extreme outburst observed in 1997.  In choosing the values for the SED model parameters, we attempted to be consistent with previous work in choosing $\gamma_{min}$, $\gamma_{max}$, $B$, $R$, and $D$ while still matching the data (see Table \ref{tab:sscmodels}).  In addition, when applying the models to the data from the two states, we attempted to limit the differences to the electron spectral indices and break energy ($\gamma_{b}$), similar to \cite{Pian} and \cite{magic501}, with a small shift in $B$ included to provide a better match to the data.  The SSC model for the 1997 outburst was matched to the data from \citet{Pian} and then the shape compared to the data from \citet{cat}.  The resulting SSC models can reproduce the measured spectra from the keV to the VHE range for both the quiescent state and the 1997 outburst.

The successful match of the SED models to the data from both the outburst and the quiescent state primarily by a modification of the injected electron spectrum implies that this may be the primary explanation for the dramatic shift in the peak frequency.  Specifically, $\gamma_{b}$ and the power-law spectral index above the break appear to drive the changes.  In addition, there is a difference in electron densities, with a density of $2.7\times10^4$ cm$^{-2}$ for the 1997 data and $2.2\times10^3$ cm$^{-2}$ for the 2009 data.  The SSC model matches indicate that during the high-flux state the break energy is shifted an order of magnitude higher and that past the break energy the spectrum is significantly harder than in the low-flux state.  It should be noted that the match of these models to the data is not unique due to the number of free parameters.  Although we did not fully explore the multi-dimensional parameter space of the model, which is not the main objective of this work, the general conclusion on the shift of the SED peaks seems to be robust.

The X-ray data do not allow us to measure the location of the low-energy SED peaks directly, but they can be used to place limits on the peak energies.  From inspection of the X-ray data points alone, the keV peak should be near 230 keV ($5.5 \times 10^{19}$ Hz) during the 1997 outburst and near 0.6 keV ($1.5 \times 10^{17}$ Hz) during the quiescent state.  This means that with a change in flux of around one order of magnitude between the quiescent state and the 1997 outburst, the keV peak of the SED shifts in frequency by more than two orders of magnitude.  

In contrast to this large shift, the VHE peak of the SED does not seem to shift as dramatically in location.  The SED peak at VHE is better constrained with the 2009 data than with the 1997 data, but the SED models applied to the data imply that even with a dramatic shift in flux at VHE, the peak energy in the VHE range is stable compared to that in the keV range.  This may be due to Klein-Nishina (KN) effects which reduce the cross-section for scattering when dealing with electrons of high energy, as also noted by \citet{Pian}.  To examine the possible KN effects, which become important above $h \nu \sim m_ec^2$ in the electron rest frame, we examined the energies of typical electrons from the model distributions.

From the 1997 SED model, the peak attributed to synchrotron radiation is located at $3.6\times 10^{19}$ Hz.  The peak energy of the emission in the jet frame, taking the Doppler factor of $D=20$ into account, is $E_{peak} \approx 7.4$ keV.  From \cite{rybicki}, it can be shown that for synchrotron radiation:

 \begin{equation}
\gamma_{e}^2 \approx \frac{8 E_{peak} m_e c}{3 \pi q B \hbar}\;,
 \label{eqn:gammae}
\end{equation}

\noindent where $q$ is charge, $\gamma_e$ is the characteristic Lorentz factor of the electrons contributing to the bulk of the X-ray emission and $E_{peak}$ is the peak energy of the synchrotron model.  This yields a value of $\gamma_e = 1.5\times10^6(B/0.23\mbox{ G})^{-1/2}(D/20)^{-1/2}$ for the 1997 outburst.  For the 2009 model, the synchrotron emission model peaks at $2.24\times10^{17}$ Hz, giving a peak energy in the jet frame of  $E_{peak} \approx 46$ eV.  This yields a value of $\gamma_e = 1.0\times10^5(B/0.34\mbox{ G})^{-1/2}(D/20)^{-1/2}$ for the 2009 measurements.

Using these electron energies, we can examine the importance of KN effects during the course of the two measurements.  For the 1997 outburst, $\gamma_{e}h \nu_{peak} \approx 1.1\times10^4$ MeV, well above the electron rest mass energy of 0.511 MeV, placing this model scenario in the extreme KN range.  For the 2009 quiescent state model, $\gamma_{e}h \nu_{peak} \approx 4.6$ MeV, moderately above the KN limit.  The extreme energies involved during the 1997 outburst may indicate that the emission is piling up in the VHE range due to the reduced KN cross-section.  

The SED models also allow us to examine the characteristic cooling timescales of the synchrotron and inverse-Compton processes that are assumed to be the sources of the keV and VHE emission, respectively.  The synchrotron cooling time is shown by \cite{rybicki} to be:

 \begin{equation}
\tau_{syn} \approx \frac{6\pi m_e c}{\sigma_{T} \gamma_{e} B^2}\;,
 \label{eqn:syn}
\end{equation}

\noindent where $\sigma_T$ is the Thompson cross-section and $\gamma_e$ is the characteristic Lorentz factor of the electrons contributing to the bulk of the X-ray emission.  The inverse-Compton cooling timescale was calculated for the models, taking into account KN effects using the method discussed in \citet{kncalc}. As discussed above, models for both sets of observations indicate that KN effects are important, so these effects were carefully included in the calculation. Using the parameters for the 1997 model, the synchrotron cooling timescale is comparable to the SSC cooling timescale ($\tau_{syn} \approx 9.8 \times 10^3$ s, $\tau_{SSC} \approx 3.7\times 10^3$ s) for electrons with $\gamma_e = 1.5\times10^6(B/0.23\mbox{ G})^{-1/2}(D/20)^{-1/2}$. However, for the quiescent state model from 2009, the timescale for inverse-Compton cooling is two orders of magnitude shorter than the synchrotron cooling timescale ($\tau_{syn} \approx 6.7 \times 10^4$ s, $\tau_{SSC} \approx 4.6 \times 10^2$ s) for electrons with $\gamma_e = 1.0\times10^5(B/0.34\mbox{ G})^{-1/2}(D/20)^{-1/2}$, indicating that in this model scenario, radiative cooling is dominated by the SSC process for the values of $\gamma_e$ calculated above.  This seems to indicate that the SSC peak energy is determined by the transition into the KN regime rather than the maximum or peak electron energies.  Beyond this transition energy, the ratio of inverse-Compton to synchrotron cooling timescales should decrease dramatically.  An estimate for the KN transition can be found using the peak of the synchrotron spectrum.  For example, from the 2009 data, this peak resides at 46 eV in the jet frame.  This corresponds to a value of $\epsilon'= E'/(m_e c^2)=9\times10^{-5}$.  The transition to the KN regime is then at the observed energy of $E_{KN} \sim m_e c^2 D (1/\epsilon') \sim 110$ GeV, very near the observed SSC peak.  In addition, all of the timescales calculated were comparable to or less than the light-crossing time for the modeled emission regions, indicating that light travel time effects may dominate the light curve profiles.

Further intense multiwavelength observations of the source, in conjunction with long-term monitoring campaigns, will continue to be important to understand the broadband behavior of Markarian 501 and to shed light on the blazar phenomenon in general.

\bigskip % extra skip inserted
\begin{acknowledgments}
{\it Acknowledgments.  }We would like to thank Elena Pian for providing the BeppoSAX data.  This work was supported in part by NASA through grants NNX08AZ98G and
NNX08AX53G.  This research has made use of data obtained from the Suzaku satellite, a collaborative mission between the space agencies of Japan (JAXA) and the USA (NASA).  The VERITAS research is supported by grants from the U.S. Department of Energy, the U.S. National Science Foundation and the Smithsonian Institution, by NSERC in Canada, by Science Foundation Ireland and by STFC in the U.K.  We acknowledge the excellent work of the technical support staff at the Fred Lawrence Whipple Observatory and the collaborating institutions in the construction and operation of the instrument.

The \textit{Fermi}-LAT Collaboration acknowledges generous ongoing support
from a number of agencies and institutes that have supported both the
development and the operation of the LAT as well as scientific data analysis.
These include the National Aeronautics and Space Administration and the
Department of Energy in the United States, the Commissariat \`a l'Energie Atomique
and the Centre National de la Recherche Scientifique / Institut National de Physique
Nucl\'eaire et de Physique des Particules in France, the Agenzia Spaziale Italiana
and the Istituto Nazionale di Fisica Nucleare in Italy, the Ministry of Education,
Culture, Sports, Science and Technology (MEXT), High Energy Accelerator Research
Organization (KEK) and Japan Aerospace Exploration Agency (JAXA) in Japan, and
the K.~A.~Wallenberg Foundation, the Swedish Research Council and the
Swedish National Space Board in Sweden.  Additional support for science analysis during the operations phase is gratefully
acknowledged from the Istituto Nazionale di Astrofisica in Italy and the Centre National d'\'Etudes Spatiales in France.

The MAGIC collaboration would like to thank the Instituto de Astrof\'{\i}sica de
Canarias for the excellent working conditions at the
Observatorio del Roque de los Muchachos in La Palma.
The support of the German BMBF and MPG, the Italian INFN, 
the Swiss National Fund SNF, and the Spanish MICINN is 
gratefully acknowledged. This work was also supported by 
the Marie Curie program, by the CPAN CSD2007-00042 and MultiDark
CSD2009-00064 projects of the Spanish Consolider-Ingenio 2010
programme, by grant DO02-353 of the Bulgarian NSF, by grant 127740 of 
the Academy of Finland, by the YIP of the Helmholtz Gemeinschaft, 
by the DFG Cluster of Excellence ``Origin and Structure of the 
Universe'', and by the Polish MNiSzW Grant N N203 390834.

\end{acknowledgments}

%{\it Facilities:} \facility{FermiLAT}, \facility{MAGIC}, \facility{VERITAS}, \facility{Suzaku}

\bigskip % extra skip inserted
\newpage

\newpage

\begin{table}[t]
\begin{center}
\scriptsize
\caption{Comparisons of SED models for Mrk501 in different flux states from previous work.  The models from this work are shown in the last two rows.  The 1997 high-flux-state data from \citet{Pian} were re-matched for this work with the same SSC code that we used to match the data from the 2009 low-flux state (See Sec. \ref{sec:spectrum} for a description of the model parameters).  The models are plotted in Figure \ref{fig:sed}. \newline}

\begin{tabular}{ l  c  c  c  c  c  c  c  c }
\hline \hline {Data Set} & {$\gamma_{min}$} & {$\gamma_{max}$} & {$\gamma_{b}$} & {$p_1$}& {$p_2$} & {B (G)}  & {R (cm)} & {D}\\ \hline
Albert et al. (2007, Low flux) & 1& $5\times10^6$ & $10^5$& 2&3.2&0.31&$1\times10^{15}$&25\\
Anderhub et al. (2009, Low flux) & 1&$1\times10^7$&$6\times10^4$&2&3.9&0.313&$1.03\times10^{15}$&20\\
Petry et al. (2000, Low flux) & 500&$3\times10^7$&-&2.4&-&0.05&$3\times10^{15}$&30\\
Pian et al. (1998, Low flux) & $3\times10^3$&$6\times10^5$&-&2&-&0.8&$5\times10^{15}$&15\\
Pian et al. (1998, High flux)& $4\times10^5$&$3\times10^6$&-&1&-&0.8&$5\times10^{15}$&15\\
Sambruna et al. (2000, Low flux) & 1 &$3.9\times10^6$&-& &-&0.03&$4\times10^{16}$&25\\
This work, 2009 Low-flux state&1&$3\times10^6$&$7\times10^4$&1.4&3.6&0.34&$1\times10^{15}$&20\\
This work, 1997 High-flux state$^{a}$&1&$3\times10^6$&$3\times10^5$&1.6&2.6&0.23&$1\times10^{15}$&20\\
\hline
\end{tabular}

\footnotetext{}{$^a$Matched to data from Pian et al. (1998)}
\label{tab:sscmodels}
\end{center}
\end{table}

\end{document}